\documentclass[twocolumn,showpacs,superscriptaddress]{revtex4}
\usepackage{amssymb}
\usepackage{amsmath}
\usepackage{graphicx}
\usepackage{epsfig}
\usepackage{amsfonts}

\begin{document}

\title{ Quantum anti-Zeno effect without rotating wave approximation}
\author{Qing Ai}
\affiliation{Institute of Theoretical Physics, Chinese Academy of
Sciences, Beijing, 100190, China}
\author{Yong Li}
\affiliation{Department of Physics and Center of Theoretical and
Computational Physics, The University of Hong Kong, Pokfulam Road,
Hong Kong, China}
\author{Hang Zheng}
\affiliation{Department of Physics, Shanghai Jiao Tong University, Shanghai 200030, China}
\author{C. P. Sun}
\affiliation{Institute of Theoretical Physics, Chinese Academy of
Sciences, Beijing, 100190, China}
\begin{abstract}
In this paper, we systematically study the spontaneous decay
phenomenon of a two-level system under the influences of both its
environment and continuous measurements. In order to clarify some
well-established conclusions about the quantum Zeno effect (QZE) and
the quantum anti-Zeno effect (QAZE), we do not use the rotating wave
approximation (RWA) in obtaining an effective Hamiltonian. We
examine various spectral distributions by making use of our present
approach in comparison with other approaches. It is found that with
respect to a bare excited state even without the RWA, the QAZE can
still happen for some cases, e.g., the interacting spectra of
hydrogen. But for a physical excited state, which is a renormalized
dressed state of the atomic state, the QAZE disappears and only the
QZE remains. These discoveries inevitably show a transition from the
QZE to the QAZE as the measurement interval changes.

\end{abstract}
 \maketitle

\section{Introduction}

The quantum Zeno effect (QZE) is vividly described as a term ``a watched pot
never boils" in some quantum version~\cite{Misra77}. Usually it is used for
describing a class of effects in which constant monitoring of a quantum
system drastically slows down its dynamic evolution ~\cite{Khalhin68,Misra77}.
This may be a coherent transition (e.g., the Rabi oscillation~\cite%
{Scully97}) and an irreversible process as well. For instance, any unstable
state can be prevented from decay when adequate measurements are frequently
applied to the system~\cite{Joos84,Frerichs91,Sakurai94,Schulman98}. Here,
the couplings to a reservoir would induce an exponential decay if there were
no measurements.

On the other hand, it was predicted that the decay could also be
enhanced by frequent measurements observed under somewhat different
conditions, leading to the so-called quantum anti-Zeno effect
(QAZE)~\cite{Kofman00,Facchi01}. When the coupling to a surrounding
environment (a reservoir) is taken into consideration, the generic
QZE may not be attainable since the required measurement interval is
out of reach in experiments. Furthermore, under the influence of the
reservoir with some spectral distribution, the decay process could
be significantly accelerated by continuous measurements.

Recently it was recognized~\cite{Zheng08,Li09} that the theoretical
prediction~\cite{Kofman00} for the reservoir-enhanced decay phenomena may be
based on the rotating wave approximation (RWA)~\cite{Scully97}, where the
counter-rotating terms are neglected as they are high-frequency oscillating.
A quite natural question follows as whether or not the existence of the QAZE
really relies on the counter-rotating term, which is usually ignored in many
applications since it possesses high frequency oscillation in the
interaction picture.

In this paper, we will generally tackle this problem by investigating the
influence of the counter-rotating term on the QAZE. Without making the RWA,
as done in Ref.~\cite{Zheng08}, we develop a direct canonical transformation
approach~\cite{Zheng95,zhu-sun99} to derive an effective Hamiltonian. It is
equivalent to the second order perturbation approach. The obtained effective
Hamiltonian is exactly solvable since it possesses the same form as that for
the case with the RWA. Our calculation properly shows that for the
spontaneous decay there exists a transition from the QZE to the QAZE as the
measurement interval changes. In other words, with respect to the bare
excited state (the product state of the atomic excited state and the vacuum
of the reservoir) in the spontaneous decay, the counter-rotating terms are
irrelevant to the occurrence of the QAZE for some spectral structures. As
predicted, the essential difference between these approaches with and
without the RWA could be disregarded in some cases.

In addition to the spectra of hydrogen atom, we extend our research to the
general situations with different kinds of spectral structures. Our finding
shows that the QAZE seems to be universal except when some certain
requirement is met for a sub-Ohmic spectrum. Furthermore, in order to
compare with the existing research~\cite{Zheng08}, we also start from the
same unitary transformation, but choose the bare excited state, which is different from the physical excited state (the one
excited from the ground state of the original Hamiltonian) in Ref.~\cite{Zheng08}, as the initial
state. Then the QAZE is again witnessed for the cases of hydrogen's spectral
structure and others as well. The discrepancy between our result and the
former one~\cite{Zheng08} is attributed to the different choices of the
initial states.

The paper is structured as follows. In the next section, with a special
transformation, we obtain the effective Hamiltonian and thus the
modification of the atomic spontaneous decay rate due to the counter-rotating terms. Sec.~III
discusses the transition from the QAZE to the QZE for different spectra. In
Sec.~IV, with the same initial state, we start from anther transformation
and arrive at the same conclusion for the hydrogen atom as the one in the
previous section. And a brief summary is concluded in Sec.~V. Furthermore,
we prove that the survival probability of the atom in the excited state is
equivalent to the survival probability of the initial state for the
spontaneous decay in Appendix~\ref{app:appendix1}. In addition to Sec.~IV,
Appendix~\ref{app:appendix3} presents the details about the calculation of
the survival probability.

\section{Effective Hamiltonian without Rotating Wave Approximation}

We generally consider the QAZE for a two-level atom interacting with a
reservoir in vacuum in the weak coupling limit. According to A. O. Caldeira
and A. J. Leggett~\cite{Leggett83}, the reservoir weakly coupled to an open
quantum system can universally be modeled as a collection of many harmonic
oscillators with annihilation (creation) operator $b_{k}$($b_{k}^{\dagger }$%
) for $k$th mode with frequency $\omega _{k}$. Let $\sigma _{x,y,z}$ be the
Pauli operators. And
\begin{equation*}
\sigma ^{\pm }=\frac{1}{2}(\sigma _{x}\pm i\sigma _{y})
\end{equation*}%
are the raising and lowering operators for the atom with the excited state $%
|e\rangle $, the ground state $|g\rangle $ and the energy level spacing $%
\Omega $, respectively. Then the total system is described by the
Hamiltonian $H=H_{0}+H_{I}$:
\begin{align}
H_{0}& =\sum_{k}\omega _{k}b_{k}^{\dagger }b_{k}+\frac{\Omega }{2}\sigma _{z}%
\text{,}  \label{H0} \\
H_{I}& =\sum_{k}g_{k}[(b_{k}+b_{k}^{\dagger })\sigma ^{+}+h.c.]\text{.}
\label{H1}
\end{align}%
Here, we have assumed the coupling constants $g_{k}$'s to be real for
simplicity. However, we would like to say that the main result does not
change if we start from a general assumption that $g_{k}$'s are complex
numbers.

As the interaction term $H_{I}$ contains the counter-rotating terms, i.e.,
the high-frequency terms with frequencies $\pm (\omega _{k}+\Omega )$ like
\begin{equation*}
V=b_{k}^{\dagger }\sigma ^{+}e^{i(\omega _{k}+\Omega )t}+h.c.
\end{equation*}%
in the interaction picture, the Hamiltonian $H$ is not exactly solvable even
for the simple cases of single mode or single excitation. We use the
generalized version~\cite{zhu-sun99} of the Fr\"{o}hlich-Nakajima
transformation $\exp (-S)$~\cite{Frohlich,Nakajima53} to eliminate the
high-frequency terms in the effective Hamiltonian. Here,
\begin{equation}
S=\sum_{k}A_{k}(b_{k}^{\dagger }\sigma ^{+}-b_{k}\sigma ^{-})
\end{equation}
is an anti-Hermitian operator, where $A_{k}$'s remain to be determined. Up
to the second order, the effective Hamiltonian $H_{\text{eff}}=\exp
(-S)H\exp (S)$ is given as
\begin{equation}
H_{\text{eff}}=H_{0}+H_{1}+\frac{1}{2}[H_{1},S]+\frac{1}{2}[H_{I},S]+\cdots
\text{,}
\end{equation}%
Now we require $b_{k}^{\dagger }\sigma ^{+}+h.c.$ to be eliminated from the
first order term
\begin{eqnarray*}
H_{1} &=&H_{I}+[H_{0},S] \\
&=&\sum_{k}[g_{k}(b_{k}+b_{k}^{\dagger })\sigma ^{+}+A_{k}(\omega
_{k}+\Omega )b_{k}^{\dagger }\sigma ^{+}+h.c.] \\
&=&\sum_{k}g_{k}(b_{k}\sigma ^{+}+h.c.)\text{.}
\end{eqnarray*}%
The above equation gives the coefficients
\begin{equation}
A_{k}=-\frac{g_{k}}{\omega _{k}+\Omega } \text{.}
\end{equation}

Note that for a state $\left\vert \Psi \right\rangle $ which fulfills the
Schr\"{o}dinger equation before the transformation, i.e.,
\begin{equation}
H\left\vert \Psi \right\rangle =i\partial _{t}\left\vert \Psi \right\rangle
\text{,}
\end{equation}%
we can prove that the state after the transformation $\left\vert \Psi
\right\rangle ^{S}=\exp (-S)\left\vert \Psi \right\rangle $ satisfies the
Schr\"{o}dinger equation,
\begin{equation}
H_{\text{eff}}\left\vert \Psi \right\rangle ^{S}=i\partial _{t}\left\vert
\Psi \right\rangle ^{S}
\end{equation}%
with the effective Hamiltonian
\begin{equation}
H_{\text{eff}}=\sum_{k}\omega _{k}b_{k}^{\dagger }b_{k}+\frac{\Omega _{1}}{2}%
\sigma _{z}+\sum_{k}g_{k}(b_{k}\sigma ^{+}+h.c.)\text{,}
\end{equation}%
where we have omitted the high-frequency intercrossing terms such as $%
b_{k}^{\dagger }b_{k^{\prime }}^{\dagger }$ and $b_{k^{\prime }}b_{k}$, and
the modified level spacing for the atom is
\begin{equation}
\Omega _{1}=\Omega +\sum_{k}\frac{g_{k}^{2}}{\omega _{k}+\Omega }\text{.}
\end{equation}%
Here, the shift of the level spacing can be regarded as the Lamb shift, and
also called the AC stark modification in atomic physics and quantum optics.
Furthermore, in the above calculation, the term
\begin{equation}
\sum_{k}\frac{g_{k}^{2}}{\omega _{k}+\Omega }(1+b_{k}^{\dagger }b_{k})
\notag
\end{equation}%
is replaced by
\begin{equation}
\sum_{k}\frac{g_{k}^{2}}{\omega _{k}+\Omega }  \notag
\end{equation}%
since for the single excitation case its contribution results in a small
modification in the $l$'th mode $g_{l}^{2}/(\omega _{l}+\Omega )$. We remark
that, for those modes $k\neq k^{\prime }$ with smaller frequency
differences, the terms $b_{k}^{\dagger }b_{k^{\prime }}$($k\neq k^{\prime }$%
) $\ $could have larger contributions in quantum dynamics, but for some
initial states we will choose them to be of the second order. This problem
has been considered in Ref.~\cite{Zheng08}.

Let us first consider the QAZE for the spontaneous decay process where the
initial state can be chosen as $\left\vert e,\{0\}\right\rangle =\left\vert
e\right\rangle \otimes \left\vert \{0\}\right\rangle $ with the atom in the
excited state $\left\vert e\right\rangle $ and all modes of fields in the
vacuum state $\left\vert \{0\}\right\rangle =\prod_{k}\otimes \left\vert
0_{k}\right\rangle $. Due to the special unitary transformation $\exp (-S)$,
the initial states before and after the transformation are identical, i.e.,
\begin{equation}
e^{-S}\left\vert e,\{0\}\right\rangle =(I-S+\frac{1}{2}S^{2})\left\vert
e,\{0\}\right\rangle =\left\vert e,\{0\}\right\rangle \text{.}
\end{equation}%
We note that, for the generalized Fr\"{o}hlich-Nakajima transformation in
Ref.~\cite{Zheng08}, the initial state would be changed. For other cases, it
will be illustrated that the uses of changed and unchanged initial states
would result in the different conclusions about the discussions of the QAZE.

When the atom is projected onto the excited state provided that the total
system evolves from the initial state $\left\vert e,\{0\}\right\rangle $,
the survival probability is
\begin{equation}
P(t)=\left\vert x(t)\right\vert ^{2}=\text{Tr}(\left\vert e\right\rangle
\left\langle e\right\vert e^{-iHt}\left\vert e,\{0\}\right\rangle
\left\langle e,\{0\}\right\vert e^{iHt})\text{.}  \label{P}
\end{equation}%
Thus, as shown in Appendix \ref{app:appendix1}, the survival probability
after $n$ measurements
\begin{align}
P(t=n\tau )& =\left\vert \left\langle e,\{0\}\right\vert e^{-iH\tau
}\left\vert e,\{0\}\right\rangle \right\vert ^{2n}  \notag \\
& =\left\vert \left\langle e,\{0\}\right\vert e^{S}e^{-iH_{\text{eff}}\tau
}e^{-S}\left\vert e,\{0\}\right\rangle \right\vert ^{2n}  \notag \\
& =\left\vert \left\langle e,\{0\}\right\vert e^{-iH_{\text{eff}}\tau
}\left\vert e,\{0\}\right\rangle \right\vert ^{2n}  \notag \\
& =e^{-Rt}  \label{P1}
\end{align}%
is calculated in the transformed representation where the new initial state
just coincides with the original one. Here, the decay rate~\cite{Kofman00}
\begin{equation}
R=2\pi \int\nolimits_{-\infty }^{\infty }d\omega F(\omega ,\Omega
_{1})G(\omega )  \label{R}
\end{equation}%
is the overlap integral of the measurement-induced atomic level broadening
\begin{align}
F(\omega ,\Omega _{1})& =\frac{1}{2\pi }\int\nolimits_{-\infty }^{\infty
}dt(1-\frac{\left\vert t\right\vert }{\tau })e^{i2\Omega _{1}t}\theta (\tau
-\left\vert t\right\vert )e^{-i\omega t}  \notag \\
& =\frac{\tau }{2\pi }\text{sinc}^{2}[\frac{(\omega -\Omega _{1})\tau }{2}]
\label{OurF}
\end{align}%
and the interacting spectrum
\begin{equation}
G(\omega )=\int\nolimits_{-\infty }^{\infty }dt\sum_{k}\frac{g_{k}^{2}}{2\pi
}e^{i(\omega -\omega _{k})t}=\sum_{k}g_{k}^{2}\delta (\omega -\omega _{k})%
\text{.}  \label{OurG}
\end{equation}

\begin{figure}[tbp]
\includegraphics[bb=35 390 555 755,width=7
cm]{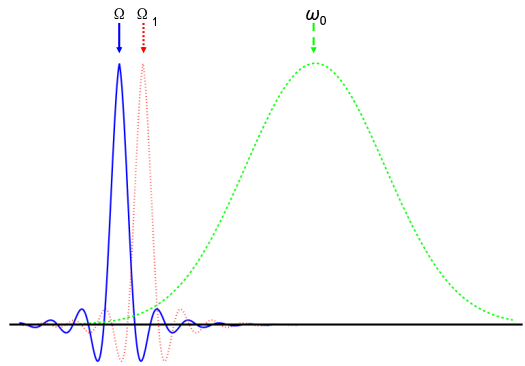}
\caption{ (color online) Schematic of the difference between the overlap
integrals with and without the RWA. The blue solid line for the measurement
function $F(\protect\omega ,\Omega )$ centered at the original atomic level
spacing $\Omega $ for the case with the RWA and the red dotted line for the
measurement function $F(\protect\omega ,\Omega _{1})$ centered at the
modified frequency $\Omega _{1}$ for our current case without the RWA, the
green dashed line for interacting spectrum $G(\protect\omega )$ centered at $%
\protect\omega _{0}$.}
\label{sketch1}
\end{figure}

The above obtained result seems to be the same as that in Ref.~\cite%
{Kofman00}, but the essential difference is that the peak of $F(\omega
,\Omega _{1})$ has been shifted due to the counter-rotating terms. In this
approach for practical problems, this shift may not have significant effect
on the physical result (see the illustration in Fig.~\ref{sketch1}). In the
physical systems that we are considering, i.e., hydrogen atom, the influence
of the counter-rotating terms is tiny small since the energy shift $%
\left\vert \Omega _{1}-\Omega \right\vert $ is relative small with respect
to the distance between the original energy level spacing $\Omega $ and the
peak of the interacting spectrum $\omega _{0}$. However, there may appear
some different results in artificial systems such as circuit QED~\cite%
{Blais04}. We will check this observation for various cases as follows.

\section{Quantum Anti-Zeno Effect for Different Interacting Spectra}

Having obtained the effective decay rate modified by the counter-rotating
terms, we examine the above observation for specific spectra in
investigating the QZE and the QAZE.

\subsection{Quantum Anti-Zeno Effect for Hydrogen Atom}

Let us first investigate the decay rate for the hydrogen atom in the vacuum
of electromagnetic fields. We consider two usual transitions of the hydrogen
atom, i.e., 2p-1s and 3p-1s, with the interacting spectra~\cite%
{Moses73,Facchi98}
\begin{equation}
G_{\text{2p-1s}}(\omega )=\frac{\eta \omega }{[1+(\frac{\omega }{\omega _{c}}%
)^{2}]^{4}}\text{,}
\end{equation}%
and
\begin{equation}
G_{\text{3p-1s}}(\omega )=\frac{\eta ^{\prime }\omega {[1+2(\frac{\omega }{%
\omega _{c}^{\prime }})^{2}]^{2}}}{[1+(\frac{\omega }{\omega _{c}^{\prime }}%
)^{2}]^{6}}\text{,}
\end{equation}%
respectively, where
\begin{align}
\eta & =6.435\ast 10^{-9}\text{,}\ \ \omega _{c}=8.491\ast 10^{18}\text{
rad/s,}  \label{eta1} \\
\eta ^{\prime }& =1.455\ast 10^{-9}\text{,}\ \ \omega _{c}^{\prime
}=7.547\ast 10^{18}\text{ rad/s.}  \label{eta2}
\end{align}

\begin{figure}[tbp]
\includegraphics[bb=0 0 270 180,width=7
cm]{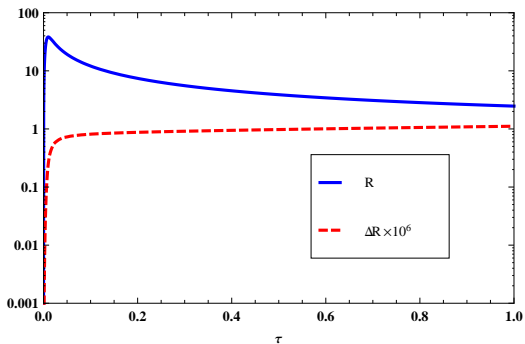}
\caption{ (color online) The decay rate vs measurement interval $\protect%
\tau $ for the 2p-1s transition of the hydrogen atom. Here, blue solid line
for $R$ and red dashed line for $\Delta R=|R-R_{\text{rwa}}|$. $|\Omega
_{1}-\Omega |/\Omega =1.71\times 10^{-6}$, $\Omega =1.55\times 10^{16}$
rad/s and $\protect\omega _{c}/\Omega =550$. Notice that $\Delta R$ is
enlarged by $10^{6}$ times. Notice that $\Delta R$ is enlarged by $10^{6}$
times. In all figures, the measurement interval $\protect\tau $ is in units
of atomic level spacing $1/\Omega $ and the decay rate $R$ is normalized
with respect to the unperturbed one $R_{0}$. }
\label{r2p1sSun}
\end{figure}

\begin{figure}[ptb]
\includegraphics[bb=0 0 270 180,width=7
cm]{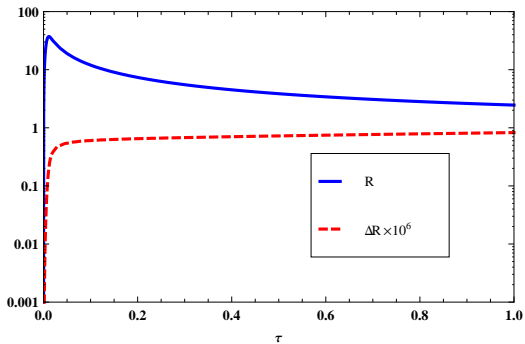}
\caption{ (color online) The decay rate vs measurement interval $\protect%
\tau $ for the 3p-1s transition of the hydrogen atom. Here, blue solid line
for $R $ and red dashed line for $\Delta R=|R-R_{\text{rwa}}|$. $%
|\Omega_{1}-\Omega|/\Omega=1.27\times10^{-6}$, $\Omega=1.83\times10^{16}$
rad/s and $\protect\omega^{\prime}_{c}/\Omega=412$. Notice that $\Delta R$
is enlarged by $10^{6}$ times. }
\label{r3p1sSun}
\end{figure}

The numerical calculations of the decay rate in Eq.~(\ref{R}) are shown in
Figs.~\ref{r2p1sSun} and \ref{r3p1sSun}. Here, we observe the emergence of
the QZE and the QAZE as well. Starting from a large enough value of $\tau $,
as the measurement interval decreases, the decay rate will experience an
ascending procedure at the first stage. Since the decay rate is bigger than
the unperturbed one%
\begin{equation*}
R_{0}=2\pi G(\Omega _{1})\text{,}
\end{equation*}%
the QAZE occurs before it reaches the climax. After the turning point, the
trend is changed. It is obvious that the decay rate drops steeply as the $%
\tau $ is further reduced. When the normalized decay rate falls below $1$,
the QZE is present. As the measurement becomes more and more frequent, i.e.,
$\tau \rightarrow 0$, we observe the transition from the QAZE to the QZE.
Mathematically speaking, the decay rate is the overlap integral of the
measurement-induced atomic level broadening $F(\omega ,\Omega )$ and the
interacting spectrum $G(\omega )$. $F(\omega ,\Omega )$ is peaked at $\Omega
$ with width $1/\tau $ while $G(\omega )$ is maximized at a frequency of the
order of the cutoff frequency $\omega _{c}$ which is much bigger than the
atomic level spacing $\Omega $. As $\tau $ decreases from a large enough
value, $F(\omega ,\Omega )$ covers more and more raising part of $G(\omega )$%
. As a consequence, the decay rate is enhanced and the QAZE is witnessed.
When the measurement interval $\tau $ is reduced to the order of $1/\omega
_{c}$, the decay rate will no longer increase since $F(\omega ,\Omega )$ has
already covered the main part of $G(\omega )$. And afterwards the opposite
phenomenon is observed. In these two figures, also shown are the differences
between the decay rates obtained from the one with the RWA and the one
without the RWA, $\Delta R=|R-R_{\text{rwa}}|$. Notice that $\Delta R$'s are
of the order of $10^{-6}$ (in units of $R_{0}$). It's a reasonable result
since the only effect of the counter-rotating terms lies in the modified
level spacing $\Omega _{1}$. And the small correction is of the order of $%
10^{-6}$ with respect to the original level spacing.

\subsection{Quantum (Anti-)Zeno Effect for General Spectral Distribution}

\begin{figure}[ptb]
\includegraphics[bb=90 590 370 770,width=7
cm]{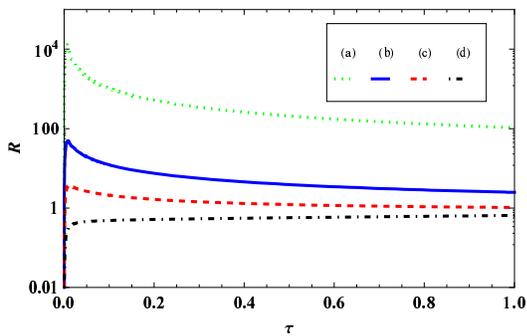}
\caption{ (color online) The decay rate vs measurement interval $\protect%
\tau $ for different spectra. Here, (a) green dotted line for super-Ohmic
spectrum $s=2$, (b) blue solid line for Ohmic spectrum $s=1$, (c) red dashed
line for sub-Ohmic spectrum $s=0.5$ and (d) black dot-dashed line for
sub-Ohmic spectrum $s=\Omega/\protect\omega_{c}$. For all spectra we set $%
A=10^{-8}$, $\protect\omega_{c}/\Omega=500$.}
\label{RdiffSpectra}
\end{figure}

Afterwards, we generally investigate the QAZE for different spectral
structures. Especially, we discover the condition when the QAZE disappears.
In general, the interacting spectra are classified as three categories. They
can be written with a uniform spectrum function~\cite{Leggett87}
\begin{equation}
G(\omega )=A\omega _{c}^{1-s}\omega ^{s}e^{-\omega /\omega _{c}}\text{,}
\label{GeneralG}
\end{equation}%
where $A$ is a constant and $\omega _{c}$ is the cutoff frequency. For an
Ohmic spectrum, $s=1$ while $s<1$ and $s>1$ for sub-Ohmic and super Ohmic
spectra respectively. In Fig.~\ref{RdiffSpectra}, the transition from the
QAZE to the QZE is again observed. It is a predictable result since the peak
of the spectrum function is located at $\omega =s\omega _{c}$. As long as $%
s\omega _{c}\gg \Omega $, the QAZE definitely occurs. Moreover, on condition
that
\begin{equation}
\Omega _{1}\simeq \Omega =s\omega _{c}\text{,}
\end{equation}%
the QAZE is wiped out and only the QZE takes place, as shown by the black
dot-dashed line in Fig.~\ref{RdiffSpectra}. Additionally, the difference
between the decay rates with and without the RWA is plotted in Fig.~\ref%
{DeltaRdiffSpectra}. Since the contribution is no more than $10^{-3}$ for
the given parameters $A=10^{-8}$ and $\omega _{c}/\Omega =500$, the
counter-rotating terms thus can be neglected as the routine work done in
quantum optics.

\begin{figure}[ptb]
\includegraphics[bb=90 585 370 770,width=7
cm]{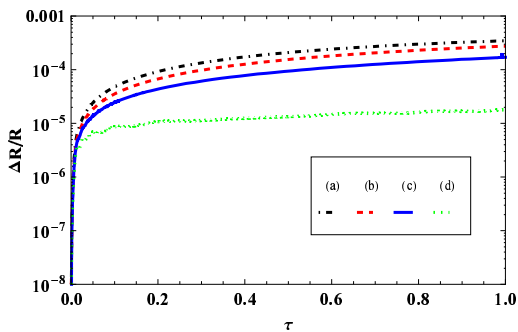}
\caption{ (color online) The decay rate difference $\Delta R=|R-R_{\text{rwa}%
}|$ vs measurement interval $\protect\tau$ for different spectra. Here, (a)
black dot-dashed line for sub-Ohmic spectrum $s=\Omega/\protect\omega_{c}$
with $|\Omega_{1}-\Omega|/\Omega=1.55\times10^{-3}$, (b) red dashed line for
sub-Ohmic spectrum $s=0.5$ with $|\Omega_{1}-\Omega|/\Omega
=9.41\times10^{-4}$, (c) blue solid line for Ohmic spectrum $s=1$ with $%
|\Omega _{1}-\Omega|/\Omega=6.57\times10^{-4}$, and (d) green dotted line
for super-Ohmic spectrum $s=2$ with $|\Omega_{1}-\Omega|/\Omega=4.00%
\times10^{-4}$. Parameters are the same as those given in Fig.~\protect\ref%
{RdiffSpectra}.}
\label{DeltaRdiffSpectra}
\end{figure}

\begin{figure}[ptb]
\includegraphics[bb=0 0 465 340,width=7
cm]{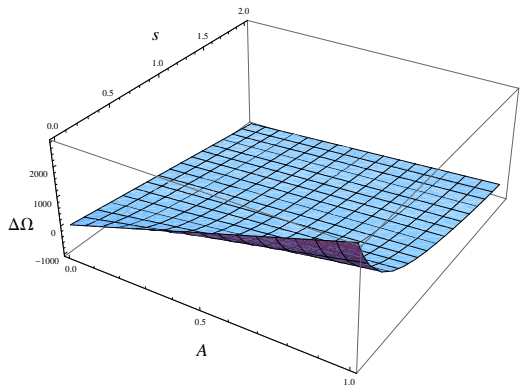}
\caption{ (color online) The frequency distance $\Delta\Omega=\Omega_1-s%
\protect\omega_c$ vs the parameters $A$ and $s$ with $\protect\omega%
_c/\Omega=500$.}
\label{NoQAZE}
\end{figure}

However, the above analysis is based on the assumption of a small energy
level shift. For some physical systems, this shift may play an important
role in the existence of the QAZE. For a given interacting spectrum as shown
in Eq.~(\ref{GeneralG}), the modified energy level spacing reads
\begin{equation}
\Omega_1=\Omega+A e^{\frac{\Omega }{\omega_c}} \Omega ^s
\Gamma(1+s,0)\Gamma(-s,\frac{\Omega }{\omega_c}) \text{,}
\label{ModifiedEnergy}
\end{equation}
where
\begin{equation}
\Gamma (u,z)=\int _z^{\infty }t^{u-1}e^{-t}dt \text{.}
\end{equation}
As stated above, the QAZE disappears if the peaks of the
measurement-induced atomic level broadening and the interacting
spectrum coincide. Therefore, the distance
$\Delta\Omega=\Omega_1-s\omega_c$ between these two peaks is plotted
vs the parameters $A$ and $s$ in Fig.~\ref{NoQAZE}. As shown, the
distance $\Delta\Omega$ increases monotonically with increasing $A$.
This is because the energy level shift, i.e., the second term on the
right hand side of Eq.~(\ref{ModifiedEnergy}), is proportional to
$A$. Physically speaking, the larger the parameter $A$ is, the
stronger the coupling between the atom and the reservoir becomes. As
a result of the stronger coupling, the energy level shift is
enlarged. Besides, it is seen that $\Delta\Omega$ falls as the
parameter $s$ raises. Thus, for a matching pair of $A$ and $s$, the
two peaks of $F(\omega,\Omega_1)$ and the interacting spectrum
$G(\omega)$ are the same. In this case, there will be only the QZE.
Besides, we also notice that the QZE was well explored for a
two-level system in either a low- or high-frequency bath beyond the
RWA \cite{Cao10}.

\section{Alternative Approach for the Decay Phenomenon of Hydrogen Atom}

In this section, with a different choice of the transformation $\exp
(-S^{\prime })$, we will obtain the same result as the preceding section for
the hydrogen atom. This transformation approach can also work well for other
spectral distributions but here we do not repeat the straightforward
calculations.

We choose the same transformation $U^{\prime }=\exp (-S^{\prime })$ as that
in Ref.~\cite{Zheng08} with
\begin{equation}
S^{\prime }=\sum_{k}A_{k}[(b_{k}^{\dagger }\sigma ^{+}-b_{k}\sigma
^{-})+(b_{k}^{\dagger }\sigma ^{-}-b_{k}\sigma ^{+})]\text{,}
\end{equation}%
and
\begin{equation}
A_{k}=\frac{-g_{k}}{\omega _{k}+\Omega }
\end{equation}%
to eliminate the counter-rotating terms $b_{k}^{\dagger }\sigma
^{+}+b_{k}\sigma ^{-}$ in the desired effective Hamiltonian. This
transformation is different from ours in that it includes the
slow-oscillating terms, i.e., $b_{k}^{\dagger }\sigma ^{-}-b_{k}\sigma ^{+}$%
. As a consequence, it modifies not only the atomic energy level spacing,
but also its coupling to the reservoir and thus the interacting spectrum.

By virtue of omitting higher order terms, i.e., $b_{k}^{\dagger
}b_{k^{\prime }}^{\dagger }$, $b_{k}b_{k^{\prime }}$, the effective
Hamiltonian is straightforward given as
\begin{equation}
H_{\text{eff}}^{\prime }=\sum_{k}\omega _{k}b_{k}^{\dagger }b_{k}+\frac{%
\Omega ^{\prime }}{2}\sigma _{z}+\sum_{k}g_{k}^{\prime }(b_{k}\sigma
^{+}+h.c.)\text{,}
\end{equation}%
where the modified coupling constant is
\begin{equation}
g_{k}^{\prime }=\frac{2\Omega }{\omega _{k}+\Omega }g_{k}\text{, }
\end{equation}%
and the modified atomic energy level spacing is
\begin{equation}
\Omega ^{\prime }=\Omega +2\sum_{k}\frac{\Omega g_{k}A_{k}}{\omega
_{k}+\Omega }
\end{equation}%
with the coefficients
\begin{equation}
A_{k}=\frac{-g_{k}}{\omega _{k}+\Omega }\text{.}
\end{equation}%
We would like to mention that the term
\begin{equation}
\sum_{k}\frac{\Omega g_{k}A_{k}}{\omega _{k}+\Omega }(1+b_{k}^{\dagger
}b_{k})  \notag
\end{equation}%
is replaced by
\begin{equation}
\sum_{k}\frac{\Omega g_{k}A_{k}}{\omega _{k}+\Omega }  \notag
\end{equation}%
in the above calculation since their contribution results in small
modification.

Since the original Hamiltonian $H$ and a state $|\psi (t)\rangle $ fulfill
the Schr\"{o}dinger equation $H|\psi (t)\rangle =i\partial _{t}|\psi
(t)\rangle $, we emphasize that it is the transformed state $|\psi
(t)\rangle ^{S^{\prime }}=\exp (-S^{\prime })|\psi (t)\rangle $ and the
effective Hamiltonian $H_{\text{eff}}^{\prime }$ that meet the same
requirement $H_{\text{eff}}^{\prime }|\psi (t)\rangle ^{S^{\prime
}}=i\partial _{t}|\psi (t)\rangle ^{S^{\prime }}$. Thus, in general cases
the initial state $|\psi (0)\rangle $ before the transformation should be
changed as $|\psi (0)\rangle ^{S^{\prime }}=\exp (-S^{\prime })|\psi
(0)\rangle $ after the transformation.

In practice, the choice of the initial state relies on the concrete physical
problem. So far as the vacuum-induced spontaneous decay is concerned, we
should choose the bare excited state. We emphasize that this choice is
consistent with the one in Ref.~\cite{Kofman00}. We also remark that it
would be more reasonable to start from the same initial state when we refer
to the influence of the counter-rotating terms on the QAZE. Besides, we can
also choose the physical excited state elsewhere, i.e., in Ref.~\cite%
{Zheng08}. It is a reasonable consideration since the ground state $%
|g,\{0\}\rangle $ of the Hamiltonian under RWA is replaced by $\exp
(S^{\prime })|g,\{0\}\rangle $ due to the presence of the counter-rotating
terms in the interaction Hamiltonian (\ref{H1})~\cite{Loudon06}. Therefore,
the initial state may be $\exp (S^{\prime })|e,\{0\}\rangle $ instead of $%
|e,\{0\}\rangle $ under the condition that the initial state is prepared
from the ground state $\exp (S^{\prime })|g,\{0\}\rangle $ through
excitation by laser. These two different choices will result in distinct
consequences.

The problem is solved in the interaction picture with respect to
\begin{equation}
U_{0}=e^{-iH_{0}^{\prime }t}
\end{equation}%
with the ``renormalized" free Hamiltonian
\begin{equation}
H_{0}^{\prime }=\sum_{k}\omega _{k}b_{k}^{\dagger }b_{k}+\frac{\Omega
^{\prime }}{2}\sigma _{z}\text{.}
\end{equation}%
And the interaction Hamiltonian $H_{1}^{I}=U_{0}H_{1}^{\prime
}U_{0}^{+}\equiv U_{0}(H_{\text{eff}}^{\prime }-H_{0}^{\prime })U_{0}^{+}$
reads
\begin{equation}
H_{1}^{I}=\sum_{k}g_{k}^{\prime }(b_{k}\sigma ^{+}e^{i(\Omega ^{\prime
}-\omega _{k})t}+b_{k}^{\dagger }\sigma ^{-}e^{-i(\Omega ^{\prime }-\omega
_{k})t})\text{.}
\end{equation}%
The time evolution of the wavefunction
\begin{equation}
\left\vert \psi _{I}^{\prime }(t)\right\rangle =\alpha (t)\left\vert
e,\{0\}\right\rangle +\sum_{k}\beta _{k}(t)\left\vert g,1_{k}\right\rangle
\label{ksai}
\end{equation}%
is governed by the effective Hamiltonian $H_{1}^{I}$,
\begin{equation}
i\partial _{t}\left\vert \psi _{I}^{\prime }\right\rangle
=H_{1}^{I}\left\vert \psi _{I}^{\prime }\right\rangle \text{.}
\end{equation}%
Here, $\left\vert g,1_{k}\right\rangle $ denotes the atom in the ground
state $\left\vert g\right\rangle $ and one excitation in the $k$th mode.

Then, the coefficients meet the following demands
\begin{align}
\dot{\alpha} &
=-i\sum_{k}g_{k}^{\prime}\beta_{k}e^{i(\Omega^{\prime}-\omega_{k})t} \text{,}
\label{alpha} \\
\dot{\beta}_{k} & =-ig_{k}^{\prime}\alpha e^{-i(\Omega^{\prime}-\omega
_{k})t} \text{.}  \label{beta}
\end{align}
However, when calculating the survival probability, we should return to the
Schr\"{o}dinger picture, i.e.,
\begin{align}
\alpha & \rightarrow\alpha e^{-i\frac{\Omega^{\prime}}{2}t} \text{,}  \notag
\\
\beta_{k} & \rightarrow\beta_{k}e^{-i(\omega_{k}-\frac{\Omega^{\prime} }{2}%
)t} \text{.}  \notag
\end{align}

We would like to remark that with the initial state $\left\vert
e,\{0\}\right\rangle $, the considered survival probability for the excited
state of the atom under the original Hamiltonian $H$ is
\begin{align}  \label{P2}
P(t)=\left\vert x(t)\right\vert ^{2}=\text{Tr}(\left\vert e\right\rangle
\left\langle e\right\vert e^{-iHt}\left\vert e,\{0\}\right\rangle
\left\langle e,\{0\}\right\vert e^{iHt})\text{.}  \notag \\
\end{align}%
Correspondingly, the effective Hamiltonian and the initial state after the
above transformation are $H_{\text{eff}}^{\prime }$ and $e^{-S^{\prime
}}\left\vert e,\{0\}\right\rangle $, respectively. Then, one has the
survival probability amplitude
\begin{widetext}
\begin{align}
x(t)    =&\left\langle e,\{0\}\right\vert
e^{S^\prime}e^{-iH^\prime_{\textrm{eff}}t}e^{-S^\prime}\left\vert
e,\{0\}\right\rangle \nonumber\\
  \simeq & C_{1}\left\langle e,\{0\}\right\vert
e^{-iH^\prime_{\textrm{eff}}t}\left\vert e,\{0\}\right\rangle
-C_{2}\sum_{k}A_{k}\left\langle e,\{0\}\right\vert
e^{-iH^\prime_{\textrm{eff}}t}\left\vert g,1_{k}\right\rangle
-C_{2}\sum_{k}A_{k}\left\langle g,1_{k}\right\vert
e^{-iH^\prime_{\textrm{eff}}t}\left\vert
e,\{0\}\right\rangle \nonumber\\
&  +\sum_{k}A_{k}^{2}\left\langle g,1_{k}\right\vert
e^{-iH^\prime_{\textrm{eff}}t}\left\vert g,1_{k}\right\rangle
\text{,} \label{xx}
\end{align}
\end{widetext}where we have dropped the off-diagonal terms for the fourth
term on the right hand side and
\begin{align}
C_{1}& =(1-\frac{1}{2}\sum_{k}A_{k}^{2})^{2}\simeq 1-\sum_{k}A_{k}^{2}\text{,%
} \\
C_{2}& =1-\frac{1}{2}\sum_{k}A_{k}^{2}\text{.}
\end{align}
In Appendix \ref{app:appendix1}, we present the detailed derivation in
obtaining the first line of Eq.~(\ref{xx}) from Eq.~(\ref{P2}). After a
series of deductions, the survival probability after $n$ measurements is
written as
\begin{equation}
P(t=n\tau )=\left\vert x(\tau )\right\vert ^{2n}=e^{-Rt}\text{,}
\end{equation}%
where the decay rate
\begin{equation}
R=2\pi \int\nolimits_{-\infty }^{\infty }d\omega F(\omega ,\Omega ^{\prime
})G^{\prime }(\omega )
\end{equation}%
is the overlap integral of the measurement-induced atomic level broadening
\begin{equation}
F(\omega ,\Omega ^{\prime })=\frac{\tau }{2\pi }\text{sinc}^{2}(\frac{\omega
-\Omega ^{\prime }}{2}\tau )
\end{equation}%
and the interacting spectrum
\begin{equation}
G^{\prime }(\omega )=\sum_{k}f(\omega _{k})g_{k}^{2}\delta (\omega -\omega
_{k})
\end{equation}%
with
\begin{equation}
f(\omega _{k})=1+\frac{(3\Omega -\Omega ^{\prime }+2\omega _{k})(\Omega
-\Omega ^{\prime })}{(\omega _{k}+\Omega )^{2}}\text{.}  \label{f}
\end{equation}%
Notice that the measurement-induced atomic level broadening differs from
ours in Eq.~(\ref{OurF}) in that it is centered at a different modified
level spacing $\Omega ^{\prime }$. And the interacting spectrum is also
altered with an additional factor $f(\omega )$ in contrary to the unaltered
one in Eq.~(\ref{OurG}). For the necessary details, please refer to Appendix~%
\ref{app:appendix3}.

\begin{figure}[ptb]
\includegraphics[bb=85 590 355 765,width=7
cm]{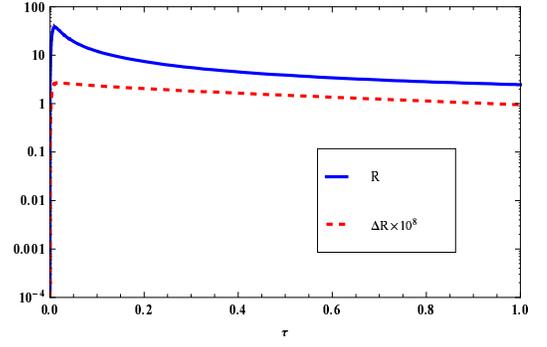}
\caption{ (color online) The decay rate vs measurement interval $\protect%
\tau $ for the 2p-1s transition of the hydrogen atom. Here, blue solid line
for $R $ and red dashed line for $\Delta R=|R-R_{\text{rwa}}|$. $|\Omega
^{\prime}-\Omega|/\Omega=5.69\times10^{-8}$ and $\protect\omega%
_{c}/\Omega=550$. Notice that $\Delta R$ is enlarged by $10^{8}$ times. }
\label{r2p1s}
\end{figure}

\begin{figure}[tbp]
\includegraphics[bb=85 555 415 770,width=7
cm]{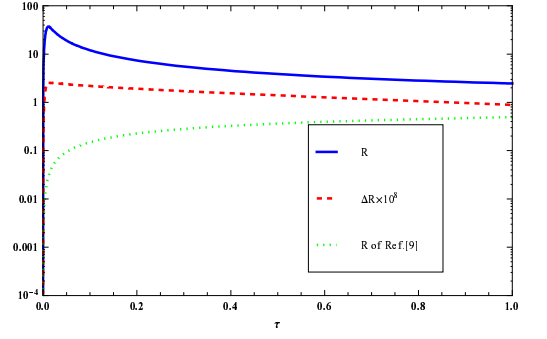}
\caption{ (color online) The decay rate vs measurement interval $\protect%
\tau $ for the 3p-1s transition of the hydrogen atom. Here, blue solid line
for $R$, red dashed line for $\Delta R=|R-R_{\text{rwa}}|$ and green dotted
line for the result from Ref.~\protect\cite{Zheng08}. $|\Omega ^{\prime
}-\Omega |/\Omega =5.32\times 10^{-8}$ and $\protect\omega _{c}/\Omega =412$%
. Notice that $\Delta R$ is enlarged by $10^{8}$ times. }
\label{r3p1s}
\end{figure}

For the hydrogen atom, the modified interacting spectrum of the 2p-1s
transition is
\begin{equation}
G_{\text{2p-1s}}^{\prime }(\omega )=f(\omega )\frac{\eta \omega }{[1+(\frac{%
\omega }{\omega _{c}})^{2}]^{4}}  \label{G2p1s}
\end{equation}%
with $\eta $ and $\omega _{c}$ already given in Eq.~(\ref{eta1}). Notice
that the second term in Eq.~(\ref{f}) is a small correction to the one with
the RWA of the order of $10^{-8}$. This is in consistence with the orders of
the numerical results of $\Delta R$ in Fig.~\ref{r2p1s}-\ref{r3p1s}.

The relation between the decay rate $R$ and the measurement interval $\tau $
is plotted in Fig.~\ref{r2p1s}. As shown in this figure, the decay rate is
well separated into two parts with the climax being the boundary. In the
left part, as the measurements are done more and more frequently, i.e., $%
\tau \rightarrow 0$, the decay rate falls monotonously. When it is less than
the decay rate without measurement%
\begin{equation*}
R_{0}^{\prime }=2\pi G^{\prime }(\Omega ^{\prime })\text{,}
\end{equation*}%
i.e., $R/R_{0}^{\prime }<1$, the QZE takes place. To the right of the
climax, the opposite trend is witnessed. In this region, the shorter the
measurement interval is, the larger the decay rate is. Since $%
R/R_{0}^{\prime }>1$ for the whole region to the right of the climax, one
can observe the QAZE, which was predicted to be obliterated due to the
counter-rotating terms~\cite{Zheng08}. The discrepancy between their result
and ours is attributed to the different choices of the initial states~\cite%
{Zheng08}. Here, we also emphasize that our approach is very simple and
concise in contrast to theirs, which is shown in Appendix~\ref{app:appendix3}%
.

For more evidence, we resort to the 3p-1s transition of the hydrogen atom.
The interacting spectrum is adjusted as
\begin{equation}
G_{\text{3p-1s}}^{\prime }(\omega )=f(\omega )\frac{\eta ^{\prime }\omega {%
[1+2(\frac{\omega }{\omega _{c}^{\prime }})^{2}]^{2}}}{[1+(\frac{\omega }{%
\omega _{c}^{\prime }})^{2}]^{6}}\text{,}  \label{G3p1s}
\end{equation}%
where $\eta ^{\prime }$ and $\omega _{c}^{\prime }$ are given in Eq.~(\ref%
{eta2}). The result of the numerical calculations is displayed in Fig.~\ref%
{r3p1s}. Here, we again observe the complete opposite predictions of the
QAZE in contrary to Ref.~\cite{Zheng08}. Moreover, the RWA offers a good
approximation in the weak-coupling limit since the disagreement between its
and the exact result is trivial.

\section{Conclusion}

In summary, we have studied the role of the counter-rotating terms
of the atomic couplings to the reservoir in the irreversible atomic
transition during the continuous measurements. By the generalized
Fr\"{o}hlich-Nakajima transformation, the exactly solvable
Hamiltonian is effectively obtained without the RWA in the form of
the standard ``spin-boson" model. We discovered that when we
consider the spontaneous decay of the bare excited state even
without RWA, the QAZE remains if the proper measurement interval is
given. And a transition from the QZE to the QAZE as the measurement
interval changes. As for the findings in Ref.~\cite{Zheng08} and its
following papers, it is observed that the disappearance of the QAZE
under the approach without RWA is mainly due to the choice of the
physical excited state. In Ref.~\cite{Zheng08}, this initial state
is the excited state of the renormalized Hamiltonian, which is
essentially an entangled state of photons and atomic states. For the
physical systems in realistic world, the influence of the
weak-coupling counter-rotating terms on the decay rate is tiny small
and can be negligible. We have utilized two different approaches for
the generalization of the Fr\"{o}hlich-Nakajima transformation. For
the same initial state, i.e., the bare excited state, consistent
conclusion is obtained for the interacting spectra of the hydrogen
atom. By comparing the two effective Hamiltonians, we find out that
in our approach there are only one parameter modified in contrary to
one more set in Ref.~\cite{Zheng08}. To further verify the
universality of the presence of the QAZE, we also extend our
investigation to different types of spectra. It is discovered that
when the cutoff frequency and the atomic level spacing fulfill some
condition, only will the QZE emerge for the sub-Ohmic spectrum. We
notice that by means of the QZE in the dynamic version a quantum
switch was proposed to control the transport of a single photon in a
one-dimensional waveguide under the RWA \cite{Zhou09}. However, we
might look forward to some novel features if no RWA is invoked.

Besides, it is worth underlining that the choice of the different
initial states depends on the specific physical problem. So far as
the QAZE for the vacuum-induced spontaneous decay is concerned, we
should choose the bare excited state in that it is the vacuum that
induces the spontaneous decay of the atomic excitation. On the other
hand, on account of the preparation of the initial state, the
physical excited state might be a better choice as well because it
can be feasibly excited from the ground state of the original
Hamiltonian.

This work was supported by NSFC through grants 10974209 and 10935010 and by
the National 973 program (Grant No.~2006CB921205).

\appendix

\section{Survival Probability $P(t)$}

\label{app:appendix1}

When we refer to the QZE and QAZE, we take a projective measurement on the
atom. Thus, we shall trace over all the possible states of the fields. Based
on the above considerations, we give the detailed deduction about the
survival probability $P(t)=\left\vert x(t)\right\vert ^{2}$.

Before the transformation, the original Hamiltonian is
\begin{equation}
H=\sum_{k}\omega _{k}b_{k}^{\dagger }b_{k}+\frac{\Omega }{2}\sigma
_{z}+\sum_{k}g_{k}(b_{k}+b_{k}^{\dagger })(\sigma ^{+}+\sigma ^{-})
\end{equation}%
with the chosen initial state to be%
\begin{equation}
\left\vert \Psi (0)\right\rangle =\left\vert e,\{0\}\right\rangle \text{.}
\end{equation}%
Then, we take a unitary transformation $e^{-S}$ with%
\begin{equation}
S=\sum_{k}A_{k}(b_{k}^{\dagger }\sigma ^{+}-b_{k}\sigma ^{-})\text{,}
\end{equation}%
the Hamiltonian is approximated to the second order as%
\begin{eqnarray}
H_{\text{eff}} &=&e^{-S}He^{S}  \notag \\
&=&\sum_{k}\omega _{k}b_{k}^{\dagger }b_{k}+\frac{\Omega _{1}}{2}\sigma
_{z}+\sum_{k}g_{k}(b_{k}\sigma ^{+}+h.c.)\text{,}  \notag \\
&&
\end{eqnarray}%
in company with a transformed initial state%
\begin{eqnarray}
\left\vert \Psi (0)\right\rangle ^{S} &=&e^{-S}\left\vert \Psi
(0)\right\rangle  \notag \\
&=&e^{-S}\left\vert e,\{0\}\right\rangle  \notag \\
&=&\left\vert e,\{0\}\right\rangle \text{.}
\end{eqnarray}%
As a result, the evolution of the state reads%
\begin{equation}
\left\vert \Psi (t)\right\rangle ^{S}=e^{-iH_{\text{eff}}t}\left\vert \Psi
(0)\right\rangle ^{S}\text{.}
\end{equation}

When calculating the survival probability for the atom in the excited state,
we shall return to the original picture and the density matrix for the total
system is straightforward given as%
\begin{eqnarray}
\rho (t) &=&\left\vert \Psi (t)\right\rangle \left\langle \Psi
(t)\right\vert   \notag \\
&=&e^{S}\left\vert \Psi (t)\right\rangle ^{SS}\left\langle \Psi
(t)\right\vert e^{-S}  \notag \\
&=&e^{S}e^{-iH_{\text{eff}}t}e^{-S}\left\vert \Psi (0)\right\rangle
\left\langle \Psi (0)\right\vert e^{S}e^{iH_{\text{eff}}t}e^{-S}  \notag \\
&=&e^{S}e^{-iH_{\text{eff}}t}e^{-S}\left\vert e,\{0\}\right\rangle
\left\langle e,\{0\}\right\vert e^{S}e^{iH_{\text{eff}}t}e^{-S}\text{.}
\notag \\
&&
\end{eqnarray}%
The reduced density matrix for the atom is traced over the degrees of
fields, i.e.,
\begin{widetext}
\begin{eqnarray}
\rho _{s}(t)&=&\text{Tr}_{B}\left\vert \Psi (t)\right\rangle
\left\langle
\Psi (t)\right\vert   \notag \\
&=&\left\langle \{0\}\right\vert
e^{S}e^{-iH_{\textrm{eff}}t}e^{-S}\left\vert e,\{0\}\right\rangle
\left\langle e,\{0\}\right\vert
e^{S}e^{iH_{\textrm{eff}}t}e^{-S}\left\vert \{0\}\right\rangle   \notag \\
&&+\sum_{k}\left\langle 1_{k}\right\vert
e^{S}e^{-iH_{\textrm{eff}}t}e^{-S}\left\vert e,\{0\}\right\rangle
\left\langle e,\{0\}\right\vert
e^{S}e^{iH_{\textrm{eff}}t}e^{-S}\left\vert 1_{k}\right\rangle
   \notag \\
&&+\sum_{k,k^\prime}\left\langle 1_{k}1_{k^\prime}\right\vert
e^{S}e^{-iH_{\textrm{eff}}t}e^{-S}\left\vert e,\{0\}\right\rangle
\left\langle e,\{0\}\right\vert
e^{S}e^{iH_{\textrm{eff}}t}e^{-S}\left\vert 1_{k}1_{k^\prime}\right\rangle
+\cdots \text{.}
\end{eqnarray}
\end{widetext}Therefore, the survival probability of the excited state of
the atom is
\begin{widetext}
\begin{eqnarray}
\rho _{s}^{ee}(t) &=&\text{Tr}_{s}(\left\vert e\right\rangle \left\langle
e\right\vert \rho _{s}(t)) \notag \\
&\simeq &\left\langle e,\{0\}\right\vert
e^{S}e^{-iH_{\textrm{eff}}t}e^{-S}\left\vert e,\{0\}\right\rangle
\left\langle e,\{0\}\right\vert
e^{S}e^{iH_{\textrm{eff}}t}e^{-S}\left\vert e,\{0\}\right\rangle \notag  \\
&&+\sum_{k}\left\langle e,1_{k}\right\vert
e^{S}e^{-iH_{\textrm{eff}}t}e^{-S}\left\vert e,\{0\}\right\rangle
\left\langle e,\{0\}\right\vert
e^{S}e^{iH_{\textrm{eff}}t}e^{-S}\left\vert e,1_{k}\right\rangle
\notag \\
&&+\sum_{k,k^\prime}\left\langle e,1_{k}1_{k^\prime}\right\vert
e^{S}e^{-iH_{\textrm{eff}}t}e^{-S}\left\vert e,\{0\}\right\rangle
\left\langle e,\{0\}\right\vert
e^{S}e^{iH_{\textrm{eff}}t}e^{-S}\left\vert e,1_{k}1_{k^\prime}\right\rangle
+\cdots\text{.}   \label{rho_ee}
\end{eqnarray}
\end{widetext}

In the following deductions, we will show that multiple excitation terms can
be omitted as they will lead to small corrections to the final result. For
the case of two excitations,
\begin{eqnarray}
e^{-S}\left\vert e,1_{k}\right\rangle &\simeq &(I-S+\frac{1}{2}%
S^{2})\left\vert e,1_{k}\right\rangle  \notag \\
&=&\left\vert e,1_{k}\right\rangle +A_{k}\left\vert g,\{0\}\right\rangle -%
\frac{1}{2}\sum_{k^{\prime }}A_{k}A_{k^{\prime }}\left\vert e,1_{k^{\prime
}}\right\rangle \text{,}  \notag \\
&&
\end{eqnarray}%
is a superposition of states with the total excitation of an even number,
while
\begin{eqnarray}
e^{-S}\left\vert e,\{0\}\right\rangle &\simeq &(I-S+\frac{1}{2}%
S^{2})\left\vert e,\{0\}\right\rangle  \notag \\
&=&\left\vert e,\{0\}\right\rangle
\end{eqnarray}%
has only one excitation. On account of $H_{\text{eff}}$'s property of
conserving the total number of excitation, the second term in Eq.~(\ref%
{rho_ee}) vanishes.

For the case of three excitations,%
\begin{eqnarray}
&&e^{-S}\left\vert e,1_{k}1_{k^{\prime }}\right\rangle  \notag \\
&\simeq &(I-S+\frac{1}{2}S^{2})\left\vert e,1_{k}1_{k^{\prime }}\right\rangle
\notag \\
&=&\left\vert e,1_{k}1_{k^{\prime }}\right\rangle +A_{k}\left\vert
g,1_{k^{\prime }}\right\rangle +A_{k^{\prime }}\left\vert
g,1_{k}\right\rangle  \notag \\
&&-\frac{1}{2}\sum_{k^{\prime \prime }}(A_{k}A_{k^{\prime \prime
}}\left\vert e,1_{k^{\prime }}1_{k^{\prime \prime }}\right\rangle
+A_{k^{\prime }}A_{k^{\prime \prime }}\left\vert e,1_{k}1_{k^{\prime \prime
}}\right\rangle )\text{.}  \notag \\
&&
\end{eqnarray}%
Then, the third term on the right hand side of Eq.~(\ref{rho_ee}) equals
\begin{eqnarray}
&&\sum_{k,k^{\prime }}\left\vert \left\langle e,1_{k}1_{k^{\prime
}}\right\vert e^{S}e^{-iH_{\text{eff}}t}e^{-S}\left\vert
e,\{0\}\right\rangle \right\vert ^{2}  \notag \\
&\simeq &2\sum_{k,k^{\prime }}A_{k}^{2}\left\vert \left\langle
g,1_{k^{\prime }}\right\vert e^{-iH_{\text{eff}}t}\left\vert
e,\{0\}\right\rangle \right\vert ^{2}  \notag \\
&&+2\left\vert \sum_{k}A_{k}\left\langle g,1_{k}\right\vert e^{-iH_{\text{eff%
}}t}\left\vert e,\{0\}\right\rangle \right\vert ^{2}\text{.}
\label{1ThirdTerm}
\end{eqnarray}

In the interaction picture, the interaction Hamiltonian reads
\begin{equation}
H^{I}=\sum_{k}g_{k}(b_{k}\sigma ^{+}e^{i(\Omega _{1}-\omega
_{k})t}+b_{k}^{\dagger }\sigma ^{-}e^{-i(\Omega _{1}-\omega _{k})t})\text{.}
\end{equation}%
The time evolution of the wavefunction
\begin{equation}
\left\vert \psi _{I}\right\rangle =\alpha (t)\left\vert e,\{0\}\right\rangle
-\sum_{k}\beta _{k}(t)\left\vert g,1_{k}\right\rangle \text{.}
\end{equation}%
is governed by the Hamiltonian $H^{I}$,
\begin{equation}
i\partial _{t}\left\vert \psi _{I}\right\rangle =H^{I}\left\vert \psi
_{I}\right\rangle \text{.}
\end{equation}

Straightforward, we attain the equations for the coefficients as
\begin{align}
\dot{\alpha}& =-i\sum_{k}g_{k}\beta _{k}e^{i(\Omega _{1}-\omega _{k})t}\text{%
,}  \label{EqAlpha} \\
\dot{\beta}_{k}& =-ig_{k}\alpha e^{-i(\Omega _{1}-\omega _{k})t}\text{.}
\label{EqBeta}
\end{align}%
The first term of Eq.~(\ref{1ThirdTerm}) is equivalent to $%
2\sum_{k,k^{\prime }}A_{k}^{2}\left\vert \beta _{k^{\prime }}(t)\right\vert
^{2}$ when $\alpha (0)=1$. We can formally integrate Eq.~(\ref{EqBeta}) and
replace $\alpha (t^{\prime })$ by $1$ to have%
\begin{eqnarray}
\beta _{k}(t) &=&-i\int_{0}^{t}dt^{\prime }g_{k}\alpha (t^{\prime
})e^{-i(\Omega _{1}-\omega _{k})t^{\prime }}  \notag \\
&\simeq &-i\int_{0}^{t}dt^{\prime }g_{k}e^{-i(\Omega _{1}-\omega
_{k})t^{\prime }}  \notag \\
&=&g_{k}\frac{e^{-i(\Omega _{1}-\omega _{k})t}-1}{\Omega _{1}-\omega _{k}}
\notag \\
&=&g_{k}\frac{-2\sin ^{2}\frac{(\Omega _{1}-\omega _{k})t}{2}-i\sin (\Omega
_{1}-\omega _{k})t}{\Omega _{1}-\omega _{k}}\text{.}
\end{eqnarray}%
As a result, the first term of Eq.~(\ref{1ThirdTerm})%
\begin{eqnarray}
&&2\sum_{k,k^{\prime }}A_{k}^{2}\left\vert \beta _{k^{\prime
}}(t)\right\vert ^{2}  \notag \\
&=&2\sum_{k}A_{k}^{2}\sum_{k^{\prime }}g_{k^{\prime }}^{2}\frac{4\sin ^{4}%
\frac{(\Omega _{1}-\omega _{k^{\prime }})t}{2}+\sin ^{2}(\Omega _{1}-\omega
_{k^{\prime }})t}{(\Omega _{1}-\omega _{k^{\prime }})^{2}}
\end{eqnarray}%
can be neglected based on the following considerations. For one thing, the
first term in the second summation is proportional to $t^{4}$ in the short
time limit and thus can be omitted. For another, the factor $%
\sum_{k}A_{k}^{2}$ is a small quantity, i.e., typically of the order of $%
10^{-8}$ for the hydrogen atom.

Moreover, the second term on the right hand side of Eq.~(\ref{1ThirdTerm})
is equivalent to $2\left\vert \sum_{k}A_{k}\beta _{k}(t)\exp (-i\omega
_{k}t)\right\vert ^{2}$ where $\alpha (0)=1$ and the factor $\exp (-i\omega
_{k}t)$ is due to transformation back to the Schr\"{o}dinger picture. Here,%
\begin{eqnarray*}
&&2\left\vert \sum_{k}A_{k}\beta _{k}(t)e^{-i\omega _{k}t}\right\vert ^{2} \\
&=&2\left\vert \sum_{k}A_{k}g_{k}\frac{-2\sin ^{2}\frac{(\Omega _{1}-\omega
_{k})t}{2}-i\sin (\Omega _{1}-\omega _{k})t}{\Omega _{1}-\omega _{k}}%
\right\vert ^{2} \\
&=&\frac{t^{4}}{2}[\int_{-\infty }^{\infty }d\omega \text{sinc}^{2}\frac{%
(\Omega _{1}-\omega )t}{2}\sum_{k}\frac{\Omega _{1}-\omega }{\Omega
_{1}+\omega }g_{k}^{2}\delta (\omega -\omega _{k})]^{2} \\
&&+2t^{2}[\int_{-\infty }^{\infty }d\omega \text{sinc}(\Omega _{1}-\omega
)t\sum_{k}\frac{g_{k}^{2}\delta (\omega -\omega _{k})}{\Omega _{1}+\omega }%
]^{2}
\end{eqnarray*}%
where the first term is proportional to $t^{4}$\ in the short time limit,
and the second term is of higher order with respect to the first term in
Eq.~(\ref{rho_ee}).

As a consequence, the second and third terms on the right hand side of Eq.~(%
\ref{rho_ee}) can be neglected. In other words, the contributions from the
multiple-excitation terms result in a small correction to the final result
and thus we have
\begin{equation}
\rho _{s}^{ee}(t)=\left\vert \left\langle e,\{0\}\right\vert e^{S}e^{-iH_{%
\text{eff}}t}e^{-S}\left\vert e,\{0\}\right\rangle \right\vert ^{2}\text{,}
\label{rho_ee2}
\end{equation}%
which is exactly the same as the one in Eq.~(\ref{P1}) for a single
measurement.

For the second approach in Sec. IV with
\begin{equation}
S^{\prime }=\sum_{k}A_{k}[(b_{k}^{\dagger }\sigma ^{+}-b_{k}\sigma
^{-})+(b_{k}^{\dagger }\sigma ^{-}-b_{k}\sigma ^{+})]\text{,}
\end{equation}%
we still have Eq.~(\ref{rho_ee}),
\begin{widetext}
\begin{eqnarray}
\rho _{s}^{ee}(t) &=&\text{Tr}_{s}(\left\vert e\right\rangle \left\langle
e\right\vert \rho _{s}(t)) \notag \\
&\simeq &\left\langle e,\{0\}\right\vert
e^{S^\prime}e^{-iH^\prime_{\textrm{eff}}t}e^{-S^\prime}\left\vert e,\{0\}\right\rangle
\left\langle e,\{0\}\right\vert
e^{S^\prime}e^{iH^\prime_{\textrm{eff}}t}e^{-S^\prime}\left\vert e,\{0\}\right\rangle \notag  \\
&&+\sum_{k}\left\langle e,1_{k}\right\vert
e^{S^\prime}e^{-iH^\prime_{\textrm{eff}}t}e^{-S^\prime}\left\vert e,\{0\}\right\rangle
\left\langle e,\{0\}\right\vert
e^{S^\prime}e^{iH_{\textrm{eff}}t}e^{-S^\prime}\left\vert e,1_{k}\right\rangle
\notag \\
&&+\sum_{k,k^\prime}\left\langle e,1_{k}1_{k^\prime}\right\vert
e^{S^\prime}e^{-iH^\prime_{\textrm{eff}}t}e^{-S^\prime}\left\vert e,\{0\}\right\rangle
\left\langle e,\{0\}\right\vert
e^{S^\prime}e^{iH^\prime_{\textrm{eff}}t}e^{-S^\prime}\left\vert e,1_{k}1_{k^\prime}\right\rangle
+\cdots\text{.} \label{rho_eeP}
\end{eqnarray}
\end{widetext}In this case, the effective Hamiltonian is replaced by
\begin{eqnarray}
H_{\text{eff}}^{\prime } &=&e^{-S^{\prime }}He^{S^{\prime }}  \notag \\
&=&\sum_{k}\omega _{k}b_{k}^{\dagger }b_{k}+\frac{\Omega ^{\prime }}{2}%
\sigma _{z}+\sum_{k}g_{k}^{\prime }(b_{k}\sigma ^{+}+h.c.)\text{,}  \notag \\
&&
\end{eqnarray}%
while the transformed initial state is given as
\begin{eqnarray}
\left\vert \Psi (0)\right\rangle ^{S^{\prime }} &=&e^{-S^{\prime
}}\left\vert \Psi (0)\right\rangle  \notag \\
&=&e^{-S^{\prime }}\left\vert e,\{0\}\right\rangle  \notag \\
&\simeq &[I-S^{\prime }+\frac{1}{2}(S^{\prime })^{2}]\left\vert
e,\{0\}\right\rangle  \notag \\
&=&(1-\frac{1}{2}\sum_{k}A_{k}^{2})\left\vert e,\{0\}\right\rangle
-\sum_{k}A_{k}\left\vert g,1_{k}\right\rangle  \notag \\
&&+\frac{1}{2}\sum_{k,k^{\prime }}A_{k}A_{k^{\prime }}\left\vert
e,1_{k}1_{k^{\prime }}\right\rangle \text{.}
\end{eqnarray}%
Then, we obtain the transformed state at time $t$,%
\begin{equation}
\left\vert \Psi ^{\prime }(t)\right\rangle ^{S^{\prime }}=e^{-iH_{\text{eff}%
}^{\prime }t}\left\vert \Psi ^{\prime }(0)\right\rangle ^{S^{\prime }}\text{.%
}
\end{equation}%
As%
\begin{eqnarray}
e^{-S^{\prime }}\left\vert e,1_{k}\right\rangle &\simeq &[I-S^{\prime }+%
\frac{1}{2}(S^{\prime })^{2}]\left\vert e,1_{k}\right\rangle  \notag \\
&=&(1-\frac{1}{2}\sum_{k^{\prime }}A_{k^{\prime }}^{2})\left\vert
e,1_{k}\right\rangle +\sum_{k}A_{k}\left\vert g,\{0\}\right\rangle  \notag \\
&&-\sum_{k}A_{k^{\prime }}\left\vert g,1_{k}1_{k^{\prime }}\right\rangle
-\sum_{k^{\prime }}A_{k}A_{k^{\prime }}\left\vert e,1_{k^{\prime
}}\right\rangle  \notag \\
&&+\frac{1}{2}\sum_{k^{\prime },k^{\prime \prime }}A_{k^{\prime
}}A_{k^{\prime \prime }}\left\vert e,1_{k}1_{k^{\prime }}1_{k^{\prime \prime
}}\right\rangle
\end{eqnarray}%
and%
\begin{eqnarray}
e^{-S^{\prime }}\left\vert e,\{0\}\right\rangle &\simeq &[I-S^{\prime }+%
\frac{1}{2}(S^{\prime })^{2}]\left\vert e,\{0\}\right\rangle  \notag \\
&=&(1-\frac{1}{2}\sum_{k^{\prime }}A_{k^{\prime }}^{2})\left\vert
e,0\right\rangle -\sum_{k}A_{k}\left\vert g,1_{k}\right\rangle  \notag \\
&&+\frac{1}{2}\sum_{k^{\prime },k}A_{k^{\prime }}A_{k}\left\vert
e,1_{k}1_{k^{\prime }}\right\rangle
\end{eqnarray}%
are of even and odd numbers of total excitations respectively, we have a
vanishing second term on the right hand side of Eq.~(\ref{rho_eeP}).

Because the contribution from%
\begin{eqnarray}
&&e^{-S^{\prime }}\left\vert e,1_{k}1_{k^{\prime }}\right\rangle  \notag \\
&\simeq &[I-S^{\prime }+\frac{1}{2}(S^{\prime })^{2}]\left\vert
e,1_{k}1_{k^{\prime }}\right\rangle  \notag \\
&=&\left\vert e,1_{k}1_{k^{\prime }}\right\rangle +A_{k}\left\vert
g,1_{k^{\prime }}\right\rangle +A_{k^{\prime }}\left\vert
g,1_{k}\right\rangle  \notag \\
&&-\sum_{k^{\prime \prime }}A_{k^{\prime \prime }}\left\vert
g,1_{k}1_{k^{\prime }}1_{k^{\prime \prime }}\right\rangle -\frac{1}{2}%
\sum_{k^{\prime \prime }}A_{k}A_{k^{\prime \prime }}\left\vert
e,1_{k^{\prime }}1_{k^{\prime \prime }}\right\rangle  \notag \\
&&+A_{k}A_{k^{\prime }}\left\vert e,0\right\rangle +\frac{1}{2}%
\sum_{k^{\prime \prime },k^{\prime \prime \prime }}A_{k^{\prime \prime
}}A_{k^{\prime \prime \prime }}\left\vert e,1_{k}1_{k^{\prime }}1_{k^{\prime
\prime }}1_{k^{\prime \prime \prime }}\right\rangle  \notag \\
&&-\frac{1}{2}\sum_{k^{\prime \prime }}A_{k}A_{k^{\prime \prime }}\left\vert
e,1_{k^{\prime }}1_{k^{\prime \prime }}\right\rangle -\frac{1}{2}%
\sum_{k^{\prime \prime }}A_{k^{\prime }}A_{k^{\prime \prime }}\left\vert
e,1_{k}1_{k^{\prime \prime }}\right\rangle  \notag \\
&&-\frac{1}{2}\sum_{k^{\prime \prime }}A_{k^{\prime }}A_{k^{\prime \prime
}}\left\vert e,1_{k}1_{k^{\prime \prime }}\right\rangle -\frac{1}{2}%
\sum_{k^{\prime \prime }}A_{k^{\prime \prime }}^{2}\left\vert
e,1_{k}1_{k^{\prime }}\right\rangle  \notag \\
\end{eqnarray}%
is of the order higher than $A_{k}^{2}$, we can further omit the third term
of Eq.~(\ref{rho_eeP}).

In total, we can neglect all nonzero-photon terms in Eq.~(\ref{rho_eeP}).
Thus, based on the above calculations, we still have
\begin{align}
\rho _{s}^{ee}(t)& =\left\vert x(t)\right\vert ^{2}=\left\vert \left\langle
e,\{0\}\right\vert e^{S^{\prime }}e^{-iH_{\text{eff}}^{\prime
}t}e^{-S^{\prime }}\left\vert e,\{0\}\right\rangle \right\vert ^{2}  \notag
\\
&
\end{align}%
for the second approach, where the survival probability $x(t)$ is the same
as Eq.~(\ref{xx}).

Judging from the above reductions, we may safely arrive at the conclusion
that for both cases the survival probability of the atom in the excited
state coincides with the survival probability of the initial state $%
|e,\{0\}\rangle $ and the single excitation approximation is reasonable.

\section{Survival Probability Amplitude $x(t)$ for the Second Approach}

\label{app:appendix3}

As shown in Eq.~(\ref{xx}), the survival probability $x(t)$ is the summation
of four terms due the modified initial state after the transformation. Since
the calculation of the first term in Eq.~(\ref{xx}) was already shown
elsewhere, i.e., Ref.~\cite{Kofman00}, we offer the detailed calculation of
the remaining parts in addition to the first term.

For convenience, we multiply $x(t)$ [Eq.~(\ref{xx})] by a factor $\exp
(i\Omega ^{\prime }t/2)$, and have
\begin{widetext}
\begin{align}
x(t)e^{i\frac{\Omega^{\prime}}{2}t} =&C_{1}\left\langle
e,\{0\}\right\vert
e^{-iH_{\textrm{eff}}t}\left\vert e,\{0\}\right\rangle e^{i\frac{\Omega^{\prime}}{2}%
t}  -C_{2}\sum_{k}A_{k}\left\langle e,\{0\}\right\vert
e^{-iH_{\textrm{eff}}t}\left\vert
g,1_{k}\right\rangle e^{i\frac{\Omega^{\prime}}{2}t}\nonumber\\
&  -C_{2}\sum_{k}A_{k}\left\langle g,1_{k}\right\vert
e^{-iH_{\textrm{eff}}t}\left\vert e,\{0\}\right\rangle
e^{i\frac{\Omega^{\prime}}{2}t}  +\sum_{k}A_{k}^{2}\left\langle
g,1_{k}\right\vert e^{-iH_{\textrm{eff}}t}\left\vert
g,1_{k}\right\rangle e^{i\frac{\Omega^{\prime}}{2}t}\text{.} \label{x}%
\end{align}
\end{widetext}

On the right hand side of the above equation, the first term
\begin{equation}
\left\langle e,\{0\}\right\vert e^{-iH_{\text{eff}}t}\left\vert
e,\{0\}\right\rangle
\end{equation}%
is equivalent to $\alpha (t)$ when $\alpha (0)=1$. Eq.~(\ref{beta}) can be
formally integrated to yield
\begin{equation}
\beta _{k}=-i\int\nolimits_{0}^{t}dt^{\prime }g_{k}^{\prime }\alpha
e^{-i(\Omega ^{\prime }-\omega _{k})t^{\prime }}\text{.}
\end{equation}%
By substituting it into Eq.~(\ref{alpha}), we have
\begin{equation}
\dot{\alpha}=-\sum_{k}\int\nolimits_{0}^{t}dt^{\prime }(g_{k}^{\prime
})^{2}\alpha e^{-i(\Omega ^{\prime }-\omega _{k})t^{\prime }}e^{i(\Omega
^{\prime }-\omega _{k})t}\text{.}
\end{equation}%
For a sufficient short time $t$, we can replace $\alpha (t_{2})$ with $%
\alpha (0)=1$ and thus
\begin{align}
\alpha (t)& \simeq
1-\int\nolimits_{0}^{t}dt_{1}\int\nolimits_{0}^{t_{1}}dt_{2}\sum%
\nolimits_{k}(g_{k}^{\prime })^{2}e^{i(\Omega ^{\prime }-\omega
_{k})(t_{1}-t_{2})}  \notag \\
& =1-t\int\nolimits_{0}^{t}dt^{\prime }(1-\frac{t^{\prime }}{t})e^{i\Omega
^{\prime }t^{\prime }}\sum_{k}(g_{k}^{\prime })^{2}e^{-i\omega _{k}t^{\prime
}}  \notag \\
& =1-I_{\alpha }(t)\text{.}
\end{align}%
When transforming it back to the Schr\"{o}dinger picture, we have
\begin{equation}
\left\langle e,\{0\}\right\vert e^{-iH_{\text{eff}}t}\left\vert
e,\{0\}\right\rangle =\alpha e^{-i\frac{\Omega ^{\prime }}{2}t}\text{.}
\end{equation}%
By multiplying a factor $\exp (i\Omega ^{\prime }t/2)$, the time-dependent
factor is canceled,
\begin{equation}
\left\langle e,\{0\}\right\vert e^{-iH_{\text{eff}}t}\left\vert
e,\{0\}\right\rangle e^{i\frac{\Omega ^{\prime }}{2}t}=\alpha \text{.}
\end{equation}%
And the following function will be used in the calculation of $x(t)$,
\begin{align}
2\text{Re}I_{\alpha }(t)& =2\text{Re}[t\int\nolimits_{0}^{t}dt^{\prime }(1-%
\frac{t^{\prime }}{t})e^{i\Omega ^{\prime }t^{\prime
}}\sum_{k}(g_{k}^{\prime })^{2}e^{-i\omega _{k}t^{\prime }}]  \notag \\
& =2\pi t\int\nolimits_{-\infty }^{\infty }d\omega F(\omega ,\Omega ^{\prime
})G_{1}(\omega )\text{,}
\end{align}%
which is the overlap integral of the measurement function
\begin{align}
F(\omega ,\Omega ^{\prime })& =\frac{1}{2\pi }\int\nolimits_{-\infty
}^{\infty }dt^{\prime }(1-\frac{\left\vert t^{\prime }\right\vert }{t}%
)e^{i\Omega ^{\prime }t^{\prime }}\theta (t-\left\vert t^{\prime
}\right\vert )e^{-i\omega t^{\prime }}  \notag \\
& =\frac{t}{2\pi }\text{sinc}^{2}(\frac{\omega -\Omega ^{\prime }}{2}t)\text{%
,}
\end{align}%
and the interacting spectrum
\begin{align}
G_{1}(\omega )& =\frac{1}{2\pi }\int\nolimits_{-\infty }^{\infty }dt^{\prime
}\sum_{k}(g_{k}^{\prime })^{2}e^{-i\omega _{k}t^{\prime }}e^{i\omega
t^{\prime }}  \notag \\
& =\sum_{k}(g_{k}^{\prime })^{2}\delta (\omega -\omega _{k})\text{.}
\end{align}

Before calculating the second and third terms in Eq.~(\ref{x}), we prove
these two terms to be equal to simplify the calculations. For a general
Hamiltonian $H$, which is time-independent and satisfies $H_{ab}\equiv
\left\langle a\right\vert H\left\vert b\right\rangle =H_{ba}$ for any two
states $\left\vert a\right\rangle $ and $\left\vert b\right\rangle $ in the
complete Hilbert space, one has
\begin{align}
F_{ab}& =\left\langle a\right\vert e^{-iHt}\left\vert b\right\rangle  \notag
\\
& =\left\langle a\right\vert \sum_{n}\frac{\left( -itH\right) ^{n}}{n!}%
\left\vert b\right\rangle  \notag \\
& =\left\langle a\right\vert \sum_{n}\frac{\left( -itH\right) ^{2n}}{(2n)!}%
\left\vert b\right\rangle +\left\langle a\right\vert \sum_{n}\frac{\left(
-itH\right) ^{2n+1}}{(2n+1)!}\left\vert b\right\rangle  \notag \\
& \equiv \text{Re}(F_{ab})+i\text{Im}(F_{ab})\text{.}
\end{align}%
It is obvious that the first term on the right hand side of the above
equation is real and the second term is pure image. On the other hand, one
has
\begin{align}
& \left\langle b\right\vert e^{-iHt}\left\vert a\right\rangle  \notag \\
=&\left[ \left( \left\langle b\right\vert e^{-iHt}\left\vert a\right\rangle
\right) ^{\dagger }\right] ^{\dagger }  \notag \\
=&[\left\langle a\right\vert e^{iHt}\left\vert b\right\rangle ]^{\dagger }
\notag \\
=&\left[ \left\langle e\right\vert \sum_{n}\frac{\left( itH\right) ^{2n}}{%
(2n)!}\left\vert b\right\rangle +\left\langle a\right\vert \sum_{n}\frac{%
\left( itH\right) ^{2n+1}}{(2n+1)!}\left\vert b\right\rangle \right]
^{\dagger }  \notag \\
=&\left[ \left\langle a\right\vert \sum_{n}\frac{\left( -itH\right) ^{2n}}{%
(2n)!}\left\vert b\right\rangle -\left\langle a\right\vert \sum_{n}\frac{%
\left( -itH\right) ^{2n+1}}{(2n+1)!}\left\vert b\right\rangle \right]
^{\dagger }  \notag \\
=&\left[ \text{Re}(F_{ab})-i\text{Im}(F_{ab})\right] ^{\dagger }  \notag \\
\equiv&\text{Re}(F_{ab})+i\text{Im}(F_{ab})  \notag \\
\equiv&F_{ab}\text{.}
\end{align}%
The above condition $H_{ab}=H_{ba}$\ (for any two states $\left\vert
a\right\rangle $ and $\left\vert b\right\rangle $) for a general Hamiltonian
$H$, means $g_{k}=g_{k}^{\ast }$ in our current case. In short, one has
\begin{equation}
\left\langle e,\{0\}\right\vert e^{-iH_{\text{eff}}t}\left\vert
g,1_{k}\right\rangle =\left\langle g,1_{k}\right\vert e^{-iH_{\text{ eff}%
}t}\left\vert e,\{0\}\right\rangle \text{.}
\end{equation}

For the second term of Eq.~(\ref{x})
\begin{equation}
\sum_{k}A_{k}\left\langle e,\{0\}\right\vert e^{-iH_{\text{eff}}t}\left\vert
g,1_{k}\right\rangle \text{,}
\end{equation}%
it is equal to $\alpha (t)$ for $\beta _{p}(0)=\delta _{pk}$ under Eq.~(\ref%
{alpha}). Thus,
\begin{align}
\alpha & =-i\int\nolimits_{0}^{t}dt^{\prime }\sum_{p}g_{p}^{\prime }\beta
_{p}(t^{\prime })e^{i(\Omega ^{\prime }-\omega _{k})t^{\prime }}  \notag \\
& =-i\int\nolimits_{0}^{t}dt^{\prime }g_{k}^{\prime }e^{i(\Omega ^{\prime
}-\omega _{k})t^{\prime }}  \notag \\
& =-g_{k}^{\prime }\frac{e^{i(\Omega ^{\prime }-\omega _{k})t}-1}{\Omega
^{\prime }-\omega _{k}}\text{.}
\end{align}%
Here, we will also multiply a factor $\exp (i\Omega ^{\prime }t/2)$ to
remove the time dependent factor during the transformation. Therefore,
\begin{align}
4\text{Re}I_{\beta }(t)& =4\text{Re}\sum_{k}A_{k}\left\langle
e,\{0\}\right\vert e^{-iH_{\text{eff}}t}\left\vert g,1_{k}\right\rangle e^{i%
\frac{\Omega ^{\prime }}{2}t}  \notag \\
& =4\text{Re}[-i\int\nolimits_{0}^{t}dt^{\prime }e^{i\Omega ^{\prime
}t^{\prime }}\sum_{k}A_{k}g_{k}^{\prime }e^{-i\omega _{k}t^{\prime }}]
\notag \\
& =4\sum_{k}\frac{2\Omega g_{k}^{2}}{(\omega _{k}+\Omega )^{2}}\frac{\cos
(\Omega ^{\prime }-\omega _{k})t-1}{\Omega ^{\prime }-\omega _{k}}  \notag \\
& =2\pi t\int\nolimits_{0}^{\infty }F(\omega ,\Omega ^{\prime })G_{2}(\omega
)d\omega \text{,}
\end{align}%
where
\begin{equation}
G_{2}(\omega )=\sum_{k}\frac{4\Omega (\omega -\Omega ^{\prime })}{(\omega
+\Omega )^{2}}g_{k}^{2}\delta (\omega -\omega _{k})\text{.}
\end{equation}%
Here, we emphasize that the final result does not depend on the assumption
that $g_{k}=g_{k}^{\ast }$. That's because it is the real parts of the
second and third terms that contribute to the decay rate.

For the fourth term in Eq.~(\ref{x}), we can also formally integrate $\alpha
$ to have
\begin{equation}
\alpha=-i\int\nolimits_{0}^{t}dt^{\prime}\sum_{k}g_{k}^{\prime}\beta
_{k}(t^{\prime})e^{i(\Omega^{\prime}-\omega_{k})t^{\prime}} \text{,}
\end{equation}
and substitute it into Eq.~(\ref{beta}) to yield
\begin{align}
\dot{\beta}_{k} &
=-\int\nolimits_{0}^{t}dt^{\prime}g_{k}^{\prime}\sum_{k^{\prime}}g_{k^{%
\prime}}\beta_{k^{\prime}}(t^{\prime})e^{i(\Omega
^{\prime}-\omega_{k^{\prime}})t^{\prime}}e^{-i(\Omega^{\prime}-\omega_{k})t}
\notag \\
&
=-\int\nolimits_{0}^{t}dt^{\prime}e^{-i\Omega^{\prime}(t-t^{%
\prime})}(g_{k}^{\prime})^{2}e^{i\omega_{k}(t-t^{\prime})} \text{,}
\end{align}
where we have replaced
\begin{equation}
\beta_{k^{\prime}}(t^{\prime})\simeq\beta_{k^{\prime}}(0)=\delta_{k^{%
\prime}k} \text{.}
\end{equation}
By one more iteration, we have
\begin{equation}
\beta_{k}(t)
=1-\int\nolimits_{0}^{t}dt_{1}\int\nolimits_{0}^{t_{1}}dt_{2}(g_{k}^{%
\prime})^{2}e^{i(\omega_{k}-\Omega^{\prime})(t_{1}-t_{2})}\text{.}
\end{equation}

Thus,
\begin{align}
& \sum_{k}A_{k}^{2}\left\langle g,1_{k}\right\vert e^{-iH_{\text{eff}%
}t}\left\vert g,1_{k}\right\rangle e^{i\frac{\Omega ^{\prime }}{2}t}  \notag
\\
=& \sum_{k}A_{k}^{2}\beta _{k}(t)e^{-i(\omega _{k}-\frac{\Omega ^{\prime }}{2%
})t}e^{i\frac{\Omega ^{\prime }}{2}t}  \notag \\
=& \sum_{k}A_{k}^{2}e^{-i(\omega _{k}-\Omega ^{\prime })t}+I_{\gamma }(t)
\notag \\
\simeq & \sum_{k}A_{k}^{2}e^{-i(\omega _{k}-\Omega ^{\prime })t}\text{,}
\end{align}%
where%
\begin{equation}
I_{\gamma }(t)=-t\int_{0}^{t}dt^{\prime }\left( 1-\frac{t^{\prime }}{t}%
\right) \sum_{k}A_{k}^{2}(g_{k}^{\prime })^{2}e^{-i(\omega _{k}-\Omega
^{\prime })(t^{\prime }-t)}  \notag
\end{equation}%
is negligible since it is proportional to $g_{k}^{4}$.

In total,
\begin{align}
xe^{i\frac{\Omega ^{\prime }}{2}t}& =C_{1}(1-I_{\alpha })-2C_{2}I_{\beta
}+\sum_{k}A_{k}^{2}e^{-i(\omega _{k}-\Omega ^{\prime })t}  \notag \\
& \simeq 1-I_{\alpha }-2I_{\beta }+I_{\delta }\text{,}
\end{align}%
where
\begin{equation}
I_{\delta }=\sum_{k}A_{k}^{2}[e^{-i(\omega _{k}-\Omega ^{\prime })t}-1]\text{%
,}
\end{equation}%
and we have dropped higher order terms of $\sum_{k}A_{k}^{2}$. Then, the
survival probability after one measurement is
\begin{align}
P(t)& =\left\vert x(t)\right\vert ^{2}  \notag \\
& \simeq 1-2\text{Re}I_{\alpha }-4\text{Re}I_{\beta }+2\text{Re}I_{\delta }
\notag \\
& \simeq \exp (-2\text{Re}I_{\alpha }-4\text{Re}I_{\beta }+2\text{Re}%
I_{\delta })
\end{align}%
with
\begin{align}
2\text{Re}I_{\delta }(t)& =\sum_{k}\frac{-4g_{k}^{2}}{(\omega _{k}+\Omega
)^{2}}\sin ^{2}\frac{(\omega _{k}-\Omega ^{\prime })t}{2}  \notag \\
& =\sum_{k}\int\nolimits_{-\infty }^{\infty }d\omega \frac{-4g_{k}^{2}\delta
(\omega -\omega _{k})}{(\omega +\Omega )^{2}}\sin ^{2}\frac{(\omega -\Omega
^{\prime })t}{2}  \notag \\
& =-2\pi t\int\nolimits_{-\infty }^{\infty }d\omega F(\omega ,\Omega
^{\prime })G_{3}(\omega )
\end{align}%
and
\begin{equation}
G_{3}(\omega )=\sum_{k}\frac{(\omega -\Omega ^{\prime })^{2}}{(\omega
+\Omega )^{2}}g_{k}^{2}\delta (\omega -\omega _{k})\text{.}
\end{equation}

Straightforward, the survival probability after $n$ continuous measurements
is
\begin{equation}
P(t) =\left\vert x(\tau)\right\vert ^{2n} = e^{-Rt} \text{,}
\end{equation}
where there are three contributions to the total decay rate
\begin{equation}
R=R_{1}+R_{2}+R_{3} \text{,}
\end{equation}
namely
\begin{align}
R_{1} & =2n\text{Re} I_{\alpha}(\tau)/t  \notag \\
& =2\pi\int\nolimits_{-\infty}^{\infty}d\omega
F(\omega,\Omega^\prime)G_{1}(\omega) \text{,} \\
R_{2} & =4n\text{Re} I_{\beta}(\tau)/t  \notag \\
& =2\pi\int\nolimits_{-\infty}^{\infty}d\omega
F(\omega,\Omega^\prime)G_{2}(\omega) \text{,} \\
R_{3} & =2n\text{Re} I_{\delta}(\tau)/t  \notag \\
& =2\pi\int\nolimits_{-\infty}^{\infty}d\omega
F(\omega,\Omega^\prime)G_{3}(\omega) \text{.}
\end{align}

To conclude, the total decay rate is further simplified as
\begin{equation}
R=2\pi\int\nolimits_{-\infty}^{\infty}d\omega
F(\omega,\Omega^\prime)G^\prime(\omega)\text{,}
\end{equation}
where the measurement function is
\begin{equation}
F(\omega,\Omega^\prime) =\frac{\tau}{2\pi}\text{sinc}^{2}(\frac{%
\omega-\Omega^{\prime}}{2}\tau) \text{,}
\end{equation}
and the modified interacting spectrum
\begin{align}
G^\prime(\omega) & =G_{1}(\omega)+G_{2}(\omega)+G_{3}(\omega)  \notag \\
& =\sum_{k}f(\omega_k)g_{k}^{2}\delta(\omega-\omega_{k})
\end{align}
with the factor
\begin{equation}
f(\omega_k)=1+\frac{(3\Omega-\Omega^{\prime}+2\omega
_{k})(\Omega-\Omega^{\prime})}{(\omega_{k}+\Omega)^{2}} \text{.}
\end{equation}

\end{document}